\documentclass{article} %
\usepackage{preprint_conference_arxiv,times}

\usepackage{amsmath,amsfonts,bm}

\def\eqref#1{equation~\ref{#1}}

\def\1{\bm{1}}

\def\vp{{\bm{p}}}

\DeclareMathAlphabet{\mathsfit}{\encodingdefault}{\sfdefault}{m}{sl}
\SetMathAlphabet{\mathsfit}{bold}{\encodingdefault}{\sfdefault}{bx}{n}

\newcommand{\R}{\mathbb{R}}

\DeclareMathOperator*{\argmin}{arg\,min}

\usepackage{hyperref}
\usepackage{url}
\usepackage{booktabs}  %
\usepackage{graphicx}  %
\usepackage{amsmath}   %
\usepackage{bm}        %
\usepackage{siunitx}
\usepackage{enumitem}

\makeatletter
\newcommand{\captionof}[1]{\def\@captype{#1}\caption}
\makeatother

\newcommand{\vpgs}{\vp^*}                              %
\newcommand{\dens}{\rho}
\newcommand{\densgs}{\rho^*}                           %
\newcommand{\denspred}{\rho^{\mathrm{pred}}}
\newcommand{\ES}{E}                            %
\newcommand{\Egs}{E^*}                                 %

\newcommand{\Func}{F}                        %
\newcommand{\Funcgs}{F^*}                      %
\newcommand{\vecr}{\mathbf{r}}
\newcommand{\grad}{\nabla}
\newcommand{\norm}[1]{\left\lVert#1\right\rVert}

\newcommand{\modelname}{\mbox{beSEN}}
\newcommand{\todo}[1]{\textcolor{red}{\textbf{[TODO: #1]}}}
\newcommand{\citetodo}[1]{\todo{add citation}}
\DeclareMathOperator{\mean}{mean}

\title{Electron-Density-Informed Molecular Property Prediction Using Surrogate Functionals}

\author{Manuel V. Klockow, Simon Wagner, Roman Remme, Dominic Plein \& Fred A. Hamprecht\\
Interdisciplinary Center for Scientific Computing (IWR), Heidelberg University\\
69120 Heidelberg, Germany\\
}

\iclrfinalcopy %
\begin{document}

\maketitle

\begin{abstract}
Machine learning methods are increasingly used to predict chemical properties directly from the molecular structure to bypass the often prohibitive cost of quantum mechanical calculations. Recent work has shown that including the ground-state electron density as input to the machine learning model can improve accuracy in this setting. This work employs the recent surrogate functional framework originally developed for orbital-free density functional theory and extends it to predicting molecular properties. We show that passing the ground-state density as input to the model, while simultaneously enforcing it to be the minimum of the surrogate functional, surpasses state-of-the-art performance for six out of seven extensive properties of the QM9 dataset. Additionally, the approach shows improved generalization to larger molecules taken from the QM40 dataset. On all extensive properties of both datasets, our approach outperforms models that are informed by the ground-state density but do not enforce the surrogate functional constraints.
\end{abstract}
\section{Introduction}
\label{sec:introduction}

Predicting molecular properties from first principles is a central goal of computational chemistry. However, quantum-mechanical methods such as Kohn--Sham density functional theory (DFT)~\citep{kohn1965self} remain too computationally expensive for many applications. Machine learning circumvents this cost by learning the mapping from molecular structure to properties directly from reference calculations~\citep{vonlilienfeld2020exploring, musil2021physics, keith2021combining}. In particular, graph neural networks~\citep{schutt2018schnet, anstine2025aimnet2}, including equivariant architectures~\citep{schutt2021painn, batzner2022nequip, batatia2022mace, fu2025learning, aykent2025gotennet}, have achieved remarkable accuracy on this task.

A complementary approach is to incorporate information about the electronic structure of the molecule. By the first Hohenberg--Kohn theorem~\citep{hohenberg1964inhomogeneous}, the ground-state electron density uniquely determines the external potential, up to an additive constant, and thus, in principle, all ground-state properties of the system. The density therefore provides a meaningful physical descriptor for molecular property prediction, and several approaches have used it as input to learned property models~\citep{brockherde2017bypassing, bogojeski2020quantum, bogojeski2026densnet}.

Surrogate functionals~\citep{remme2026surrogate, wagner2026strong}, originally introduced in the context of machine-learned orbital-free DFT, go beyond treating the density merely as an input representation. Instead, they construct a learned functional whose minimum is attained at the ground-state density and whose value there equals the corresponding ground-state energy. This variational structure yields energy prediction accuracy competitive with state-of-the-art interatomic potentials while substantially improving extrapolation to larger systems~\citep{wagner2026strong}.

Motivated by these results, we generalize the surrogate functional approach to arbitrary scalar molecular properties. For each target property, we train a dedicated surrogate functional using the same training pipeline, while replacing the ground-state energy with the corresponding property value as the training target. At inference time, the required electron density is supplied by a separately trained network that predicts it directly from molecular structure, avoiding a computationally expensive variational optimization.

Our contributions include:
\begin{enumerate}
    \item We extend the surrogate functional framework from the electronic energy to arbitrary scalar molecular properties.
    \item On QM9~\citep{ramakrishnan2014quantum}, surrogate functionals surpass the previous state of the art on six out of seven extensive properties, with competitive accuracy on the seventh.
    \item  In a controlled comparison using a common backbone, surrogate functionals consistently outperform both geometry-only models and models that additionally take the electron density as input across all extensive properties. This advantage becomes even more pronounced in the out-of-distribution regime, when models trained exclusively on QM9 are evaluated on the larger molecules in QM40~\citep{madushanka2024qm40}.
    \item We introduce \modelname{}, a modification of eSEN~\citep{fu2025learning} with bilinear activation based on the Gram matrix, which further improves predictive performance, particularly when used within the surrogate functional framework.
\end{enumerate}

\section{Background and related work}
\label{sec:related}

\paragraph{Molecular property prediction.}
Most machine-learned models for molecular properties map nuclear charges and coordinates directly to the quantity of interest. Starting from invariant message-passing networks~\citep{schutt2018schnet, gasteiger_fast_2022, liu_spherical_2021} and progressing through increasingly expressive equivariant architectures~\citep{schutt2021painn, batzner2022nequip, batatia2022mace, tholke_equivariant_2021, le_equivariant_2022, liao_equiformer_2022, du_new_2023, an_hybrid_2023, aykent_savenet_2023, liao2024equiformerv2, fu2025learning, aykent2025gotennet}, this line of work has reduced energy errors on the QM9 benchmark~\citep{ramakrishnan2014quantum} to a few meV relative to the reference calculations. In contrast, several models make the electronic structure explicit by using the electron density as an intermediate representation. \citet{brockherde2017bypassing} learn the map from a smoothed external potential to a density expansion, and then fit a second, independent model from the predicted density to the energy. \citet{bogojeski2020quantum} reuse this two-map construction to learn a coupled-cluster correction on top of a DFT density, and \citet{shao2023machine} extend it to the one-electron reduced density matrix.
DenSNet~\citep{bogojeski2026densnet} uses an equivariant network to predict the coefficients of a flexible atom-centered Gaussian basis, while a second equivariant network consumes these coefficients together with the nuclear positions to predict the total energy. DensIP~\citep{wing2026densip} follows the same staged recipe but replaces the second network with a four-parameter physics-based intermolecular potential, in the spirit of earlier models built on machine-learned multipoles and valence-density parameters~\citep{bereau2018noncovalent}.

\paragraph{Machine-learned orbital-free DFT.}
Orbital-free density functional theory~\citep{mi2023orbital} obtains the electronic ground state by minimizing an energy functional directly with respect to the density, without ever constructing Kohn--Sham orbitals~\citep{kohn1965self}. Recent machine-learned functionals have demonstrated energy predictions with chemical accuracy for small drug-like molecules~\citep{zhang2024overcoming, remme2025stable}. Surrogate functionals~\citep{remme2026surrogate, wagner2026strong} are trained without requiring physical energies at off-equilibrium densities, which allows training from ground-state labels alone, and have been demonstrated to exhibit improved generalization compared to directly predicting the ground-state energy.

\newpage

\section{Methods}
\label{sec:methods}
\vspace{-0.8\baselineskip}
\begin{figure}[h]
\centering
\includegraphics[width=\linewidth]{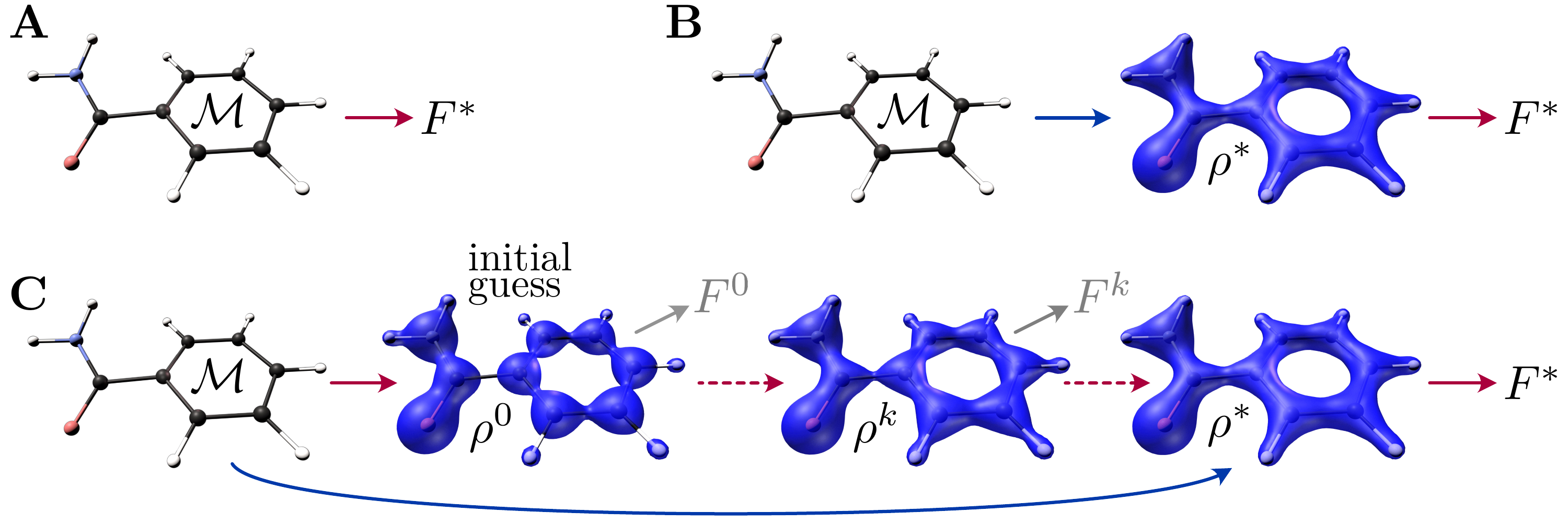}
\caption{Schematic overview of the three compared frameworks for molecular property prediction. \textbf{A} Geometry-only models predict the property directly from the molecular geometry. \textbf{B} Density-input models predict the property from the geometry in two steps via the ground-state electron density. \textbf{C} Surrogate functional models predict the property from the geometry and the ground-state density, while enforcing that the density is the minimum of a learned surrogate functional. This variational structure makes it possible to reach the ground-state density and property from an initial guess (magenta arrows). At inference time, we use the shortcut of predicting the ground-state density directly with a second network (blue arrow).}
\label{fig:schematic_overview}
\end{figure}

\vspace{-1.0\baselineskip}
\subsection{Density representation}
\label{sec:density_representation}

Following recent work in machine-learned OF-DFT~\citep{zhang2024overcoming, remme2025stable, remme2026surrogate}, we choose to express the electron density as a linear expansion in atom-centered Gaussian-type basis functions~\citep{grisafi2018transferable, vergara2023efficient}, to which symmetric L\"owdin orthonormalization~\citep{loewdin1950nonorthogonal} is applied, yielding
\begin{equation}
    \dens(\vecr) = \sum_{\mu} p_{\mu}\,\omega_{\mu}(\vecr)\,,
    \label{eq:density_expansion}
\end{equation}
with density coefficients $p_\mu$ and orthonormal basis functions $\omega_\mu(\vecr)$. We denote by $\vpgs$ the coefficients of the ground-state density $\densgs$. Each basis function is associated with a definite angular momentum $l$, so that the coefficients of an atom transform as a direct sum of irreducible representations of $\mathrm{SO}(3)$ and can be used directly as node features of an equivariant message-passing network.

\vspace{-0.2\baselineskip}
\subsection{Orbital-free surrogate functionals}
\label{sec:surrogate_functionals}

An orbital-free density functional $E[\dens]$ maps a given electron density $\dens$ onto a corresponding energy $E$. The ground-state values for both the energy and the density can be obtained by minimizing the functional over $\dens$, which in the expansion of Equation~\ref{eq:density_expansion} amounts to minimizing $E(\vp)$ over the density coefficients
\begin{equation}
    \vpgs = \argmin_{\vp} E(\vp),
    \qquad
    \Egs = E(\vpgs).
    \label{eq:functional_def}
\end{equation}
Machine-learned density functionals~\citep{zhang2024overcoming, remme2025stable} learn an approximation to parts of the physical functional from Kohn--Sham reference data. In contrast, a surrogate functional $\ES(\vp)$~\citep{remme2026surrogate, wagner2026strong} is not required to reproduce physical energies at off-equilibrium densities but only needs to converge to the correct minimum during the iterative variational optimization procedure, where Equation~\ref{eq:functional_def} is solved via gradient descent.
To ensure a stable convergence during inference, surrogate functionals are trained with a train-time variational optimization scheme: For each molecule $\mathcal{M}$, a cached coefficient vector $\vp_{\mathcal{M}}$ is maintained and updated each time the molecule appears in a batch via a gradient descent step
\begin{equation}
    \vp_{\mathcal{M}} \mapsto \vp_{\mathcal{M}} - \gamma\,\grad_\vp \ES(\vp_{\mathcal{M}})\,,
    \label{eq:cache_update}
\end{equation}
where the step size $\gamma$ is typically chosen to minimize the distance to $\vpgs$ along the predicted gradient direction~\citep{wagner2026strong}. Samples whose predicted gradient points away from the ground state are not updated, so that difficult densities remain in the cache until the model learns to improve its gradients at these points.
The central loss component is the gradient-descent-improvement (GDI) loss~\citep{remme2026surrogate}, which requires that every optimization step with step size $\lambda$ moves the coefficients closer to the true ground state by at least a predefined contraction factor $0<\beta<1$. With $\vp' = \vp - \lambda\grad_\vp\ES(\vp)$ and $[x]_+ = \max(0,x)$, the loss is
\begin{equation}
    \mathcal{L}_{\text{GDI}}
    = \left[\norm{\vp' - \vpgs} - \beta\norm{\vp - \vpgs}\right]_+\,.
    \label{eq:gdi}
\end{equation}
To also enforce the correct energy value at the minimum, the energy-bound loss~\citep{wagner2026strong} penalizes predictions that leave empirical quadratic bounds,
\begin{equation}
    \mathcal{L}_{\text{EB}}
    = \left[\Egs + a_{\text{min}}\norm{\vp - \vpgs}^2 - \ES(\vp)\right]_+
    + \left[\ES(\vp) - \Egs - a_{\text{max}}\norm{\vp - \vpgs}^2\right]_+\,.
    \label{eq:eb_loss}
\end{equation}
Furthermore, a fraction $0\leq q<1$ of each training batch is evaluated at the exact ground-state coefficients and penalized with $L_1$ losses on the predicted energy, $\mathcal{L}_{E} = |\ES(\vpgs) - \Egs|$, and on its gradient, $\mathcal{L}_{\grad E} = \norm{\grad_\vp\ES(\vpgs)}_1$, which should vanish at the desired stationary point.

\subsection{Surrogate functionals for molecular properties}
\label{sec:property_functionals}

In orbital-free DFT, the functional is the total electronic energy, and its minimization corresponds to the physically motivated variational principle. The construction itself, however, can in principle be imposed on any scalar property. \citet{wagner2026strong} demonstrated that this formulation shows superior generalization compared to methods that learn the energy directly from molecular geometry. This motivates us to also model other scalar molecular properties $P$ using surrogate functionals $\Func_P(\vp)$. For training, we use the same pipeline as described in Section~\ref{sec:surrogate_functionals} and simply replace the ground-state value of the functional with the label of the target property, $\Egs \rightarrow \Funcgs_P$. In the following, we omit the subscript $P$ whenever the property is clear from context. A schematic overview of this method compared to geometry-only and density-input models is shown in Figure~\ref{fig:schematic_overview}. We provide further information on the functional parametrization and the target scaling in Appendix \ref{app:normalization}.

\vspace{-0.2\baselineskip}
\subsection{Architecture}
\label{sec:architecture}
\vspace{-0.8\baselineskip}

\begin{figure}[h]
\centering
\includegraphics[width=\linewidth]{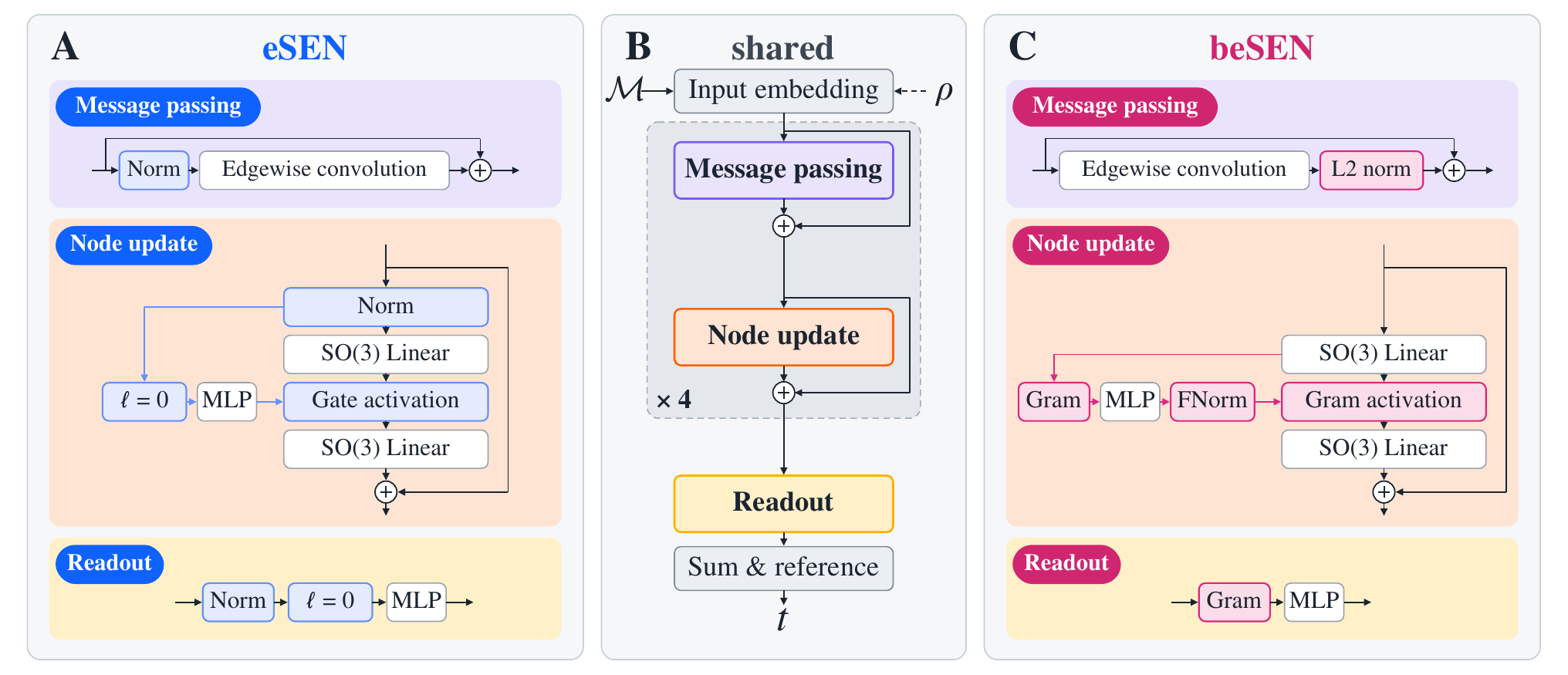}
\caption{Direct comparison of the eSEN and \modelname{} architectures. The two architectures share the same overall structure, but differ in multiple details of the message passing, node update and readout. The \modelname{} architecture is used for both the density prediction and the property prediction models, yielding superior results in both of these settings.}
\label{fig:architecture}
\end{figure}

In this section, we introduce \modelname{}, an equivariant message-passing neural network architecture based on eSEN~\citep{fu2025learning} and inspired by components of BOA~\citep{klockow2026function}. This architecture performs especially well in combination with the surrogate functional framework, as can be seen in Table~\ref{tab:backbone}. While the overall structure of the model remains very similar to eSEN, specific components of the message passing, node update and readout are changed. A direct comparison of the two architectures is shown in Figure~\ref{fig:architecture}, and the newly introduced components are described in the following subsections.

\subsubsection{Embedding}
The initial embedding is shared by the eSEN and \modelname{} architectures. It can be split into two parts: the first consumes the molecular geometry and the second consumes the density coefficients. The geometry embedding is the standard eSEN embedding~\citep{fu2025learning}. The density embedding is a simple equivariant linear layer that maps the density coefficients to the feature dimension of the model, exactly as described in \citet{wagner2026strong}. The two embeddings are then summed to produce the initial node features. 
\subsubsection{Message Passing}
The message passing step differs between the two architectures: eSEN uses a pre-norm residual connection, whereas \modelname{} uses a post-norm residual connection~\citep{xiong2020layer}. Additionally, \modelname{} uses a novel \emph{L2 norm} layer, which computes the mean per-channel L2 norm
\begin{equation}
    \mu_n = \mean_a \sqrt{\sum_p x_{apn}^2}
\end{equation}
for the first batch and then freezes this value to normalize the node features during training and inference. The resulting \(\mu_n\) is clamped at a minimum of \num{1e-6}. An additional learned per-channel weight \(\gamma_n\) is used to scale the normalized features:
\begin{equation}
    \tilde{x}_{apn} = \gamma_n \frac{x_{apn}}{\mu_n}.
\end{equation}
This normalization layer is closely related to the normalization used in BOA~\citep{klockow2026function}, but without the use of an explicit metric.

\subsubsection{Node update}
The node update of the eSEN and \modelname{} architectures differ in two places. First, \modelname{} omits the pre-norm in the node update. Second, \modelname{} uses a \emph{Gram activation} instead of the standard \emph{Gate activation}~\citep{weiler20183d}. The motivation behind both is similar: scalar features are used to gate the tensorial features. In contrast to the \emph{Gate activation}, which normally uses the scalar part of the current node features, the \emph{Gram activation} builds a Gram matrix
\begin{equation}
    G^a_{mn} = \sum_p x_{apm} x_{apn}
\end{equation}
of the feature channels and uses these scalars to gate the tensorial features. A subset of this Gram matrix is then chosen by only considering the first \(N_{\text{gate}}\) channels along the first dimension, i.e. restricting the matrix to \(\mathbf{\tilde{G}^a} \in \mathbb{R}^{N_{\text{gate}} \times N_c}\) with \(\tilde{G}^a_{mn} = G^a_{mn}\) for \(m \in \{1, \dots, N_{\text{gate}}\}\) and \(n \in \{1, \dots, N_c\}\), which is flattened and then passed through an MLP to produce the gating matrix
\begin{equation}
    \mathbf{W}^a = \operatorname{MLP}\bigl(\tilde{\mathbf{G}}^a\bigr),
  \qquad \operatorname{MLP}\colon \R^{N_{\text{gate}} \times N_c} \to \R^{N_c \times N_c}.
\end{equation}
This gating matrix \(\mathbf{W}^a \in \mathbb{R}^{N_c \times N_c}\) is then passed through a \emph{Frobenius normalization} layer
\begin{equation}
    \mu = \mean_a \sqrt{\sum_{mn} {\left(W^a_{mn}\right)^2}}, \qquad \tilde{W}^a_{mn} = \gamma \frac{W^a_{mn}}{\mu}.
\end{equation}
Similar to the \emph{L2 norm} layer, the resulting \(\mu\) is clamped at a minimum of \num{1e-6}, and an additional learned weight \(\gamma\) is used to scale the normalized gating matrix. The final node feature update is then 
\begin{equation}
    \tilde{x}_{apn} = \sum_m \tilde{W}^a_{nm} x_{apm}.
\end{equation}

\newpage

\subsubsection{Readout}
The change of the \modelname{} readout compared to the eSEN readout is similar in spirit to the change in the node update. Instead of taking the scalar features as input for a readout MLP, as in eSEN, \modelname{} uses the Gram matrix \(\mathbf{G}^a\) of the node features as input to the readout MLP. 

\subsection{Speed comparison}
\begin{figure}[htbp]
\centering
\begin{minipage}[c]{0.46\linewidth}
\centering
\captionof{table}{Forward-pass throughput on QM9, \modelname{} against eSEN at the training batch size of 128, on one idle NVIDIA A100 40GB. Median over 100 distinct full batches drawn from the training split. \modelname{} has 6.9\,M parameters against eSEN's 2.7\,M.}
\label{tab:timing}
\begin{tabular}{lrr}
\toprule
Model type & \modelname{} & eSEN \\
 & ms/batch & ms/batch \\
\midrule
Geometry only & \textbf{38.4} & 44.9 \\
Density input & \textbf{39.2} & 47.3 \\
\bottomrule
\end{tabular}
\end{minipage}%
\hfill
\begin{minipage}[c]{0.50\linewidth}
\centering
\includegraphics{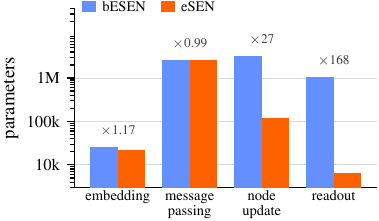}
\captionof{figure}{Number of parameters in the eSEN and \modelname{} architectures. While most of the parameters in the eSEN are in the message passing block, the parameters of \modelname{} are more evenly distributed between the message passing and the per-node update and readout.}
\label{fig:params}
\end{minipage}
\end{figure}

To offer a fair comparison between the eSEN and \modelname{} architectures on our training task, we measure the forward-pass throughput on the QM9 dataset to make sure that both architectures need a similar amount of time for training and inference. While \modelname{} has more than twice the number of parameters as the eSEN model used here, it is still slightly faster, as shown in Table~\ref{tab:timing}. Figure~\ref{fig:params} shows why the two are not in conflict: the models are the same size in the per-edge message passing, and \modelname{}'s extra parameters sit in the per-node stages, which are far cheaper to evaluate. The throughput is measured on one idle NVIDIA A100 40GB, with a training batch size of 128. The median over 100 distinct full batches drawn from the training split is reported.

\subsection{Predicting the ground-state electron density directly}
\label{sec:density_prediction}

Even though the surrogate functional framework in principle allows the simultaneous prediction of the ground-state density and the target property through variational minimization, this process is computationally expensive, requiring $\sim 100$ model forward calls for the property prediction of one molecule. Predicting the ground-state density coefficients (in the orthonormalized basis) directly with a second neural network and passing them to the main model yields comparable performance. This way, we reduce inference to two model forward calls and at the same time avoid the explicit orthonormalization of the density coefficients.
As architecture for the density network, we use the same backbone as described in Section~\ref{sec:architecture}. For the output head, we use an $\mathrm{SO}(3)$ equivariant MLP that maps to a tensorial feature, which matches the representation of the basis coefficients.  

\newpage

\section{Results}
\label{sec:results}
\vspace{-0.9\baselineskip}
We evaluate the performance of our model on the QM9 dataset~\citep{ramakrishnan2014quantum}, which contains 133\,885 small organic molecules with up to 9 heavy atoms (C, N, O, F). Additionally, the generalization of the models to larger molecules from the QM40 dataset~\citep{madushanka2024qm40} is evaluated.

\vspace{-0.6\baselineskip}
\subsection{QM9 dataset}
\vspace{-0.3\baselineskip}
We train multiple versions of \modelname{} on the QM9 dataset that differ in the type of input and training mode, but all use the same training, validation and test split as \citet{gasteiger2020directional}. One model per property is trained on the geometry only, without any density input. A second model per property is trained on the geometry and the ground-state density as input, but without any surrogate functional constraints. A third model per property is trained on the geometry and the ground-state density as input, with the surrogate functional framework applied. All models are trained with the hyperparameters listed in Appendix \ref{app:hyperparameters}. The results are shown in Table~\ref{tab:qm9-predicted}. On \(U_0\), \(U\), \(H\), \(G\), ZPVE, \(\alpha\) and \(C_v\) the surrogate functional model outperforms the geometry-only and density-input \modelname{} models by a large margin. For all of these properties but \(\alpha\), the surrogate functional \modelname{} also outperforms the previous state-of-the-art GotenNet-L model~\citep{aykent2025gotennet}. 
\vspace{-1.0\baselineskip}
\begin{table}[htbp]
\centering
\caption{Mean absolute error of the predicted properties on the QM9 test set. Values of prior work are reproduced from \citet{aykent2025gotennet}}
\label{tab:qm9-predicted}
\scriptsize
\setlength{\tabcolsep}{3pt}
\begin{tabular}{lccccccc|ccccc}
\toprule
                                                & \multicolumn{7}{c|}{\emph{Extensive}} & \multicolumn{5}{c}{\emph{Non-extensive}} \\
                                                &    $U_0$     &     $U$      &     $H$      &     $G$      &     ZPVE      &  $\alpha$   &                  $C_v$                  &     HOMO      &     LUMO      &      gap      &    $\mu$     & $\langle R^2\rangle$ \\
Unit                                            &     meV      &     meV      &     meV      &     meV      &      meV      &  $\text{m}a_0^3$   & $\frac{\mathrm{mcal}}{\mathrm{mol\,K}}$ &      meV      &      meV      &      meV      &      mD      &       $\text{m}a_0^2$       \\
\midrule
\multicolumn{13}{l}{\emph{Geometry only}} \\
\quad DimeNet++ \citep{gasteiger_fast_2022}     &     6.32     &     6.28     &     6.53     &     7.56     &     1.21      &     44      &                   23                    &     24.6      &     19.5      &     32.6      &     29.7     &         331          \\
\quad SphereNet \citep{liu_spherical_2021}      &     6.26     &     6.36     &     6.33     &     7.78     &     1.12      &     46      &                   22                    &     22.8      &     18.9      &     31.1      &     24.5     &         268          \\
\quad LEFTNet \citep{du_new_2023}               &      6       &      7       &      6       &      7       &     1.33      &     48      &                   23                    &      24       &      18       &      40       &      12      &         109          \\
\quad EQGAT \citep{le_equivariant_2022}         &      25      &      25      &      24      &      23      &     2.00      &     53      &                   24                    &      20       &      16       &      32       &      11      &         382          \\
\quad ET \citep{tholke_equivariant_2021}        &     6.15     &     6.38     &     6.16     &     7.62     &     1.84      &     59      &                   26                    &     20.3      &     17.5      &     36.1      &      11      &          33          \\
\quad HDGNN \citep{an_hybrid_2023}              &     8.34     &     8.12     &      10      &      11      &     1.21      &     46      &                   23                    &      18       &      16       &      32       &      17      &         342          \\
\quad Geoformer \citep{wang_geometric_2023}     &     4.43     &     4.41     &     4.39     &     6.13     &     1.28      &     40      &                   22                    &     18.4      &     15.4      &     33.8      &      10      &          28          \\
\quad Equiformer \citep{liao_equiformer_2022}   &     6.59     &     6.74     &     6.63     &     7.63     &     1.26      &     46      &                   23                    &     15.4      &     14.7      &      30       &      12      &         251          \\
\quad SaVeNet-B \citep{aykent_savenet_2023}     &     5.43     &     5.48     &     5.43     &     6.64     &     1.18      &     39      &                   23                    &     18.4      &     16.3      &     24.8      &     9.3      &          58          \\
\quad EquiformerV2 \citep{liao2024equiformerv2} &     6.17     &     6.49     &     6.22     &     7.57     &     1.47      &     47      &                   23                    &     14.4      &     13.3      &     29.0      &     9.9      &         186          \\
\quad GotenNet-S \citep{aykent2025gotennet}     &     3.71     &     3.67     &     3.70     &     5.50     &     1.09      &     33      &                   20                    &     16.9      &     13.9      &     21.2      &     7.5      &          29          \\
\quad GotenNet-B \citep{aykent2025gotennet}     &     3.43     &     3.49     &     3.44     &     5.19     &     1.09      &     32      &                   19                    &     15.2      &     13.0      &     20.5      &     7.2      &          27          \\
\quad GotenNet-L \citep{aykent2025gotennet}     &     3.37     &     3.41     &     3.30     &     4.98     &     1.08      & \textbf{28} &                   19                    & \textbf{13.4} & \textbf{12.2} & \textbf{19.8} & \textbf{6.7} &     \textbf{24}      \\
\quad \modelname                                &     5.3      &     5.4      &     5.3      &     6.7      &      1.1      &     41      &                   22                    &      17       &      15       &      33       &     8.2      &         150          \\
\midrule
\multicolumn{13}{l}{\emph{Density input}} \\
\quad \modelname                                &     4.9      &     4.8      &     4.8      &     6.0      &      1.1      &     39      &                   22                    &      16       &      15       &      34       &     7.9      &         140          \\
\midrule
\multicolumn{13}{l}{\emph{Surrogate}} \\
\quad \modelname                                & \textbf{2.8} & \textbf{2.4} & \textbf{2.7} & \textbf{4.2} & \textbf{0.91} &     30      &               \textbf{18}               &      16       &      15       &      32       &      12      &         480          \\
\bottomrule
\end{tabular}
\end{table}

\vspace{-0.7\baselineskip}
To evaluate the contribution of the \modelname{} architecture to the results achieved on the QM9 dataset, we compare the performance of the \modelname{} and eSEN architectures on the three evaluated training modes: geometry only, density input and surrogate functional. As shown in Table~\ref{tab:backbone}, the \modelname{} architecture outperforms the eSEN architecture on all three training modes, with the difference growing larger for the density input mode and largest for the surrogate functional mode. \\
We additionally run an extensive ablation study evaluating the impact of different training configurations. The results are discussed in Appendix \ref{app:ablation}.

\vspace{-0.8\baselineskip}
\begin{table}[htbp]
\centering
\caption{Backbone comparison on $U_0$ showing the mean absolute error (MAE) in meV on the QM9 test split. \modelname{} outperforms the eSEN architecture on all three training modes.}
\label{tab:backbone}
\begin{tabular}{lcc}
\toprule
Arm                                   & eSEN & \modelname \\
\midrule
Geometry only                         & 5.6 & \textbf{5.3} \\
\midrule
Density input                         & 5.8 & \textbf{4.9} \\
\midrule
Surrogate                             & 4.2 & \textbf{2.8} \\
\midrule
gain vs geometry only, density input  & 0.96$\times$ & \textbf{1.1$\times$} \\
gain vs geometry only, surrogate      & 1.3$\times$ & \textbf{1.9$\times$} \\
\bottomrule
\end{tabular}
\end{table}

A direct ground state density prediction model using the \modelname{} architecture is trained on QM9 electron densities. To evaluate the quality of the predicted densities, we compute the normalized mean absolute error 
\begin{equation}
    \text{NMAE} = \frac{\int \left|\densgs(\vecr) - \denspred(\vecr)\right| \text{d}\vecr}{\int \densgs(\vecr) \text{d}\vecr}.
\end{equation}
An NMAE of \(0.032\%\) is achieved on the QM9 test set. The density error is calculated with respect to the density label in the auxiliary basis, i.e.\ after density fitting (see Appendix \ref{app:data_generation}). This leads to a larger error compared to the original KS-DFT density. We evaluate the difference between the label before and after density fitting on 100 random QM9 molecules. The mean NMAE is \(0.59\%\) and therefore already more than an order of magnitude larger than the error of the direct density prediction model. The integrals necessary for the calculation of the NMAE are computed with \texttt{PySCF}~\citep{sun2018pyscf, sun2020recent} on a level-3 integration grid. Additionally, the impact of using the predicted density instead of the label density on the property prediction results is evaluated. The difference in these two densities is small enough that the property prediction results, both in the density input and surrogate setting, are only marginally affected, as shown in Table~\ref{tab:qm9-label-vs-predicted}.

\vspace{-0.9\baselineskip}

\begin{table}[htbp]
\centering
\caption{Mean absolute error of the predicted properties on the QM9 test set, evaluated at the label density and at the density predicted by the direct density prediction model.}
\label{tab:qm9-label-vs-predicted}
\scriptsize
\setlength{\tabcolsep}{3pt}
\begin{tabular}{lccccccc@{\quad}ccccc}
\toprule
                               & $U_0$ & $U$ & $H$ & $G$ & ZPVE & $\alpha$ &                  $C_v$                  & HOMO & LUMO & gap & $\mu$ & $\langle R^2\rangle$ \\
Unit                           &  meV  & meV & meV & meV & meV  & $\text{m}a_0^3$ & $\frac{\mathrm{mcal}}{\mathrm{mol\,K}}$ & meV  & meV  & meV &  mD   &       $\text{m}a_0^2$       \\
\midrule
\multicolumn{13}{l}{\emph{Density input}} \\
\quad at the label density     &  4.9  & 4.8 & 4.8 & 6.0 & 1.1  &    39    &                   22                    &  16  &  15  & 34  &  7.9  &         140          \\
\quad at the predicted density &  4.9  & 4.8 & 4.8 & 6.0 & 1.1  &    39    &                   22                    &  16  &  15  & 34  &  7.9  &         140          \\
\midrule
\multicolumn{13}{l}{\emph{Surrogate}} \\
\quad at the label density     &  2.7  & 2.3 & 2.6 & 4.0 & 0.91 &    30    &                   17                    &  16  &  15  & 32  &  12   &         430          \\
\quad at the predicted density &  2.8  & 2.4 & 2.7 & 4.2 & 0.91 &    30    &                   18                    &  16  &  15  & 32  &  12   &         480          \\
\bottomrule
\end{tabular}
\end{table}

\vspace{-0.5\baselineskip}
\subsection{Generalizing to larger molecules}

\vspace{-1.2\baselineskip}
\begin{table}[htbp]
\centering
\caption{Mean absolute error of the predicted properties on 600 QM40 molecules from 10 to 40 heavy atoms. The dataset is split into bins sorted by the number of heavy atoms, with width of 5 atoms. From each of these bins 100 molecules with atom types H, C, N, O and F that are neutral and with even electron count are selected. Each evaluated model was trained only on QM9. Especially for the extensive properties, using the surrogate functional approach enables better generalization to larger molecules.}
\label{tab:qm40-predicted}
\scriptsize
\setlength{\tabcolsep}{3pt}
\begin{tabular}{lccccccc|ccccc}
\toprule
                 & \multicolumn{7}{c|}{\emph{Extensive}} & \multicolumn{5}{c}{\emph{Non-extensive}} \\
                 &    $U_0$    &     $U$     &     $H$     &     $G$     &     ZPVE     &   $\alpha$    &                  $C_v$                  &     HOMO     &     LUMO     &      gap      &    $\mu$     &  $\langle R^2\rangle$  \\
Unit             &     meV     &     meV     &     meV     &     meV     &     meV      &   $\text{m}a_0^3$    & $\frac{\mathrm{mcal}}{\mathrm{mol\,K}}$ &     meV      &     meV      &      meV      &      mD      &        $\text{m}a_0^2$        \\
\midrule
\multicolumn{13}{l}{\emph{Geometry only}} \\
\quad GotenNet-S &     310     &     300     &     300     &     220     &      20      &     7300      &                  4900                   &     6300     &     540      &      ---      &     270      &   $2.7\times10^{5}$    \\
\quad GotenNet-B &     430     &     460     &     450     &     270     &      22      &     7400      &                  5400                   &     4700     &     520      &      ---      &     240      &          ---           \\
\quad \modelname &     180     &     180     &     190     &     120     &     150      &     3000      &                  2600                   &     310      &     490      &     1400      & \textbf{150} &   $2.3\times10^{5}$    \\
\midrule
\multicolumn{13}{l}{\emph{Density input}} \\
\quad \modelname &     140     &     130     &     150     &     110     &     150      &     3100      &                  3100                   &     300      &     510      & \textbf{1200} & \textbf{150} &   $2.4\times10^{5}$    \\
\midrule
\multicolumn{13}{l}{\emph{Surrogate}} \\
\quad \modelname & \textbf{57} & \textbf{52} & \textbf{49} & \textbf{53} & \textbf{4.9} & \textbf{2400} &              \textbf{170}               & \textbf{200} & \textbf{470} &     1300      & \textbf{150} & $\bm{1.5\times10^{5}}$ \\
\bottomrule
\end{tabular}
\end{table}

The capability to generalize to larger molecules is evaluated on 600 molecules from the QM40 dataset~\citep{madushanka2024qm40}, which contains molecules with 10 to 40 heavy atoms. The molecules are sampled from bins sorted by the number of heavy atoms, with a width of 5 heavy atoms. From each bin, 100 neutral molecules containing only the elements H, C, N, O and F and having an even number of electrons are selected. All evaluated models were trained only on QM9. The \modelname{} model is evaluated in all three training modes. Additionally, the publicly available GotenNet-S and GotenNet-B models~\citep{aykent2025gotennet} are evaluated. For the properties missing for the GotenNet models, no checkpoint was available. The results of the evaluation are shown in Table~\ref{tab:qm40-predicted}. For the extensive properties \(U_0\), \(U\), \(H\), \(G\), ZPVE, \(\alpha\) and \(C_v\) the surrogate functional model outperforms the geometry-only and density-input \modelname{} models by a large margin. The intensive properties HOMO and LUMO as well as  \(\langle R^2\rangle\) are also predicted best by the \modelname{} surrogate functional. The HOMO-LUMO gap is predicted best by the density-input \modelname{} model, while all three \modelname{}
variants perform equally on the dipole moment \(\mu\). 

\begin{figure}[h]
\centering
\includegraphics[width=\linewidth]{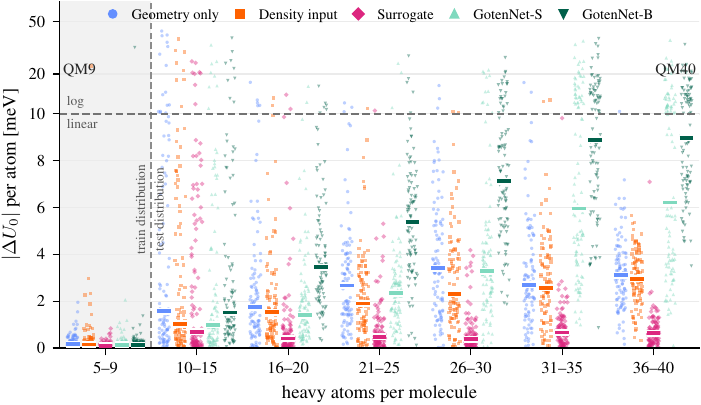}
\caption{Absolute $U_0$ error \emph{per atom} against molecule size. The surrogate and the density input models are evaluated on the density as predicted by the density prediction model. The median is shown as a horizontal line.}
\label{fig:u0-vs-size}
\end{figure}

To further illustrate the generalization capabilities of the surrogate functional model, Figure~\ref{fig:u0-vs-size} shows the absolute error per atom of the predicted \(U_0\) against the number of heavy atoms in the QM40 molecules. The surrogate functional \modelname{} shows by far the best behavior with respect to molecule size. While all other models show a clear trend of increasing error with increasing molecule size, the surrogate functional model maintains its accuracy far better, leading to much lower median errors for larger molecules. We provide a similar analysis for the remaining properties in Appendix \ref{app:size_dependence}.

\section{Discussion}
In this work, we extend the surrogate functional framework, originally developed in the context of orbital-free density functional theory, to model arbitrary scalar molecular properties. Additionally, a new equivariant message-passing neural network architecture, \modelname{}, is introduced, which performs especially well in the surrogate functional framework. Employing this framework on the QM9 dataset leads to an improvement on all extensive properties compared to the same architecture trained on geometry-only or geometry and density input. The surrogate functional framework also enables better generalization to larger molecules, as shown on the QM40 dataset.

While the surrogate functional framework has a clear motivation in the context of orbital-free density functional theory, where the electronic energy is governed by a variational principle with respect to the electron density, the motivation for using surrogate functionals for arbitrary molecular properties is less obvious. However, the substantial improvements achieved in this work suggest that the surrogate functional approach is beneficial for all extensive properties. The exact mechanism of this improvement is not yet fully understood and should be investigated in future work.

\newpage

\subsubsection*{Acknowledgments}
This work is supported by the German Research Foundation (DFG) under Germany's Excellence Strategy EXC-2181/1-390900948 (the Heidelberg STRUCTURES Excellence Cluster) as well as by the Carl-Zeiss-Stiftung via its Wildcard program.
The authors acknowledge support by the state of Baden-W\"urttemberg through bwHPC and the German Research Foundation (DFG) through grant INST 35/1597-1 FUGG.

\setlength{\bibsep}{6pt plus 0.3ex} %
\bibliography{preprint_conference}
\bibliographystyle{preprint_conference}

\newpage
\appendix
\section{Appendix}

\subsection{Ablation experiments}
\label{app:ablation}

Table~\ref{tab:ablation} summarizes ablation experiments on $U_0$ in which individual components of the surrogate training are removed or modified. All ablated configurations yield a higher error than the full setup, which combines all training components.

\vspace{-0.8\baselineskip}
\begin{table}[htbp]
\centering
\caption{Ablations of different pieces of the surrogate machinery, on $U_0$, QM9 test split. \emph{density} determines whether the model receives the predicted ground-state electron density as input. \emph{stat} is the stationarity constraint $\nabla_\vp F(\vpgs) =0$ at the ground state, \emph{bound} and \emph{GDI} are the energy-bound and GDI loss respectively and \emph{cache} says whether a cache is used. The minao column indicates how the cache is initialized at each reset, if MINAO is marked the MINAO initialization is used. Otherwise the cache entry is the ground-state density displaced along a random, electron-count-conserving direction, with the L2 displacement radius $r$ drawn log-uniformly (density $\propto 1/r$) over $[0.002, 0.3]$. \emph{gain} denotes the accuracy improvement relative to the geometry-only model.}
\label{tab:ablation}
\begin{tabular}{cccccccc}
\toprule
density & stat & bound & cache & GDI & minao & MAE & gain \\
\midrule
 &  &  &  &  &  & 5.3 & 1.00$\times$ \\
$\bullet$ &  &  &  &  &  & 4.9 & 1.09$\times$ \\
\midrule
$\bullet$ & $\bullet$ &  &  &  &  & 4.9 & 1.08$\times$ \\
$\bullet$ &  & $\bullet$ &  &  &  & 4.4 & 1.20$\times$ \\
$\bullet$ & $\bullet$ & $\bullet$ &  &  &  & 4.6 & 1.16$\times$ \\
$\bullet$ &  & $\bullet$ & $\bullet$ &  &  & 3.9 & 1.35$\times$ \\
$\bullet$ & $\bullet$ & $\bullet$ & $\bullet$ &  &  & 4.9 & 1.08$\times$ \\
\midrule
$\bullet$ &  & $\bullet$ &  &  & $\bullet$ & 4.6 & 1.15$\times$ \\
$\bullet$ & $\bullet$ & $\bullet$ &  &  & $\bullet$ & 4.6 & 1.15$\times$ \\
$\bullet$ &  & $\bullet$ & $\bullet$ &  & $\bullet$ & 4.0 & 1.34$\times$ \\
$\bullet$ & $\bullet$ & $\bullet$ & $\bullet$ &  & $\bullet$ & 5.7 & 0.93$\times$ \\
$\bullet$ &  &  & $\bullet$ & $\bullet$ & $\bullet$ & 3.3 & 1.59$\times$ \\
\midrule
$\bullet$ & $\bullet$ & $\bullet$ & $\bullet$ & $\bullet$ & $\bullet$ & \textbf{2.8} & 1.88$\times$ \\
\bottomrule
\end{tabular}
\end{table}

\newpage

\subsection{Size extrapolation for all target properties}
\label{app:size_dependence}

\begin{figure}[ht]
\centering
\includegraphics[width=\linewidth]{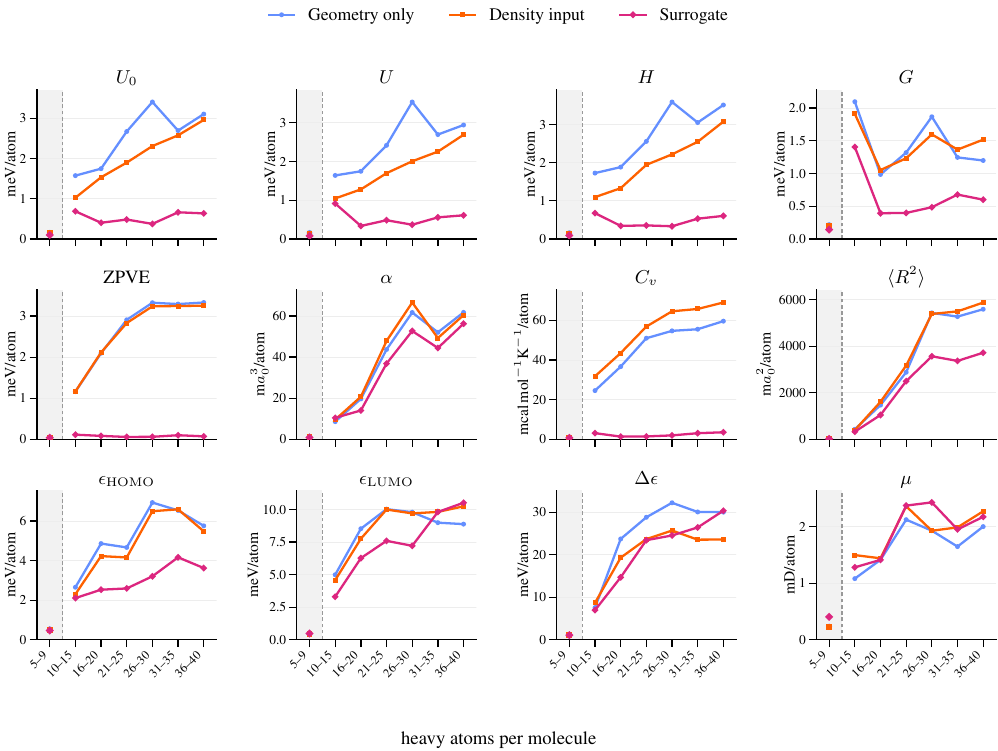}
\caption{Median absolute error per atom of the geometry-only, density-input, and surrogate \modelname{} models for all twelve target properties against the number of heavy atoms. The shaded leftmost bin corresponds to the QM9 test set and the remaining bins to the 600 QM40 molecules of Table~\ref{tab:qm40-predicted}; all models were trained on QM9 only. Errors are divided by the total number of atoms.}
\label{fig:appendix_12props}
\end{figure}

Figure~\ref{fig:appendix_12props} extends the size analysis of Figure~\ref{fig:u0-vs-size} to all twelve target properties. For the geometry-only, density-input, and surrogate \modelname{} models, it shows the median absolute error per atom on the QM9 test set and on the 600 QM40 molecules, grouped by the number of heavy atoms.

\subsection{Generation of the electron density labels}
\label{app:data_generation}

The target properties of all models are taken directly from the QM9~\citep{ramakrishnan2014quantum} and QM40~\citep{madushanka2024qm40} datasets. The ground-state electron densities are not part of these datasets and are obtained from our own Kohn--Sham DFT calculations. These are carried out with PySCF~\citep{sun2018pyscf, sun2020recent} at the B3LYP/6-31G(2df,p) level of theory that was also used to generate the QM9 dataset.

For the linear density expansion in Equation~\ref{eq:density_expansion} we use an even-tempered Gaussian basis with an exponent ratio of $2.5$. The coefficients are obtained from the Kohn--Sham densities via density fitting, using the same fitting objective as in \citet{remme2025stable}. The fitted coefficients are then orthonormalized as described in Section~\ref{sec:density_representation}.

\newpage

\subsection{Output and target scaling}
\label{app:normalization}

\subsubsection{Output parametrization}
\label{app:output_parametrization}

To ensure that the network only needs to predict quantities of order unity for every target, its output is combined with an offset and a scale that are not learned but determined once from the training set. In the following, $\mathbf{h}_a$ denotes the final node features of atom $a$ in molecule $\mathcal{M}$. Furthermore, $g_\theta$ denotes the readout network, and $\bar{F}^*$ and $\sigma_{F^*}$ denote the mean and standard deviation of the training labels $F^*$. Depending on the target property, the surrogate functional takes one of the following forms:
\begin{itemize}[leftmargin=*]
    \item \emph{Extensive properties} ($U_0$, $U$, $H$, $G$, ZPVE, $\alpha$, $C_v$):
    \begin{equation}
        \Func(\vp) = t_0 + \sum_{a \in \mathcal{M}} t_{Z_a} + \frac{1}{c} \sum_{a \in \mathcal{M}} g_\theta(\mathbf{h}_a)\,.
        \label{eq:norm_extensive}
    \end{equation}
    The atomic reference, consisting of element-specific contributions $t_Z$ and a constant term $t_0$, is obtained by linear least-squares regression of the training labels as in \citet{zhang2024overcoming}. The summed per-atom contributions are divided by a fixed normalization constant $c$, as in EquiformerV2~\citep{liao2024equiformerv2}.
    \item \emph{Intensive properties} (HOMO, LUMO, gap):
    \begin{equation}
        \Func(\vp) = \bar{F}^* + \sigma_{F^*}\, g_\theta\Bigl(\frac{1}{N_\mathcal{M}} \sum_{a \in \mathcal{M}} \mathbf{h}_a\Bigr)\,,
        \label{eq:norm_intensive}
    \end{equation}
    where $N_\mathcal{M}$ is the number of atoms. Mean-pooling the node features before the readout ensures that the prediction does not grow with molecular size.
    \item \emph{Dipole moment $\mu$ and electronic spatial extent $\langle R^2 \rangle$}: Following GotenNet~\citep{aykent2025gotennet}, we use the dedicated output heads
    \begin{equation}
        \Func_\mu(\vp) = \sigma_{F^*} \Bigl\lVert \sum_{a \in \mathcal{M}} \bigl(q_a \mathbf{R}_a + \mathbf{d}_a\bigr) \Bigr\rVert\,,
        \qquad
        \Func_{R^2}(\vp) = \sigma_{F^*} \sum_{a \in \mathcal{M}} \lVert \mathbf{R}_a - \mathbf{R}_\text{c} \rVert^2\, x_a\,,
        \label{eq:norm_heads}
    \end{equation}
    where the atomic charges $q_a$, atomic dipoles $\mathbf{d}_a$, and scalars $x_a$ are predicted from the node features $\mathbf{h}_a$, $\mathbf{R}_a$ are the atomic positions, and $\mathbf{R}_\text{c}$ is the center of mass.
\end{itemize}

\subsubsection{Target scaling}
\label{app:target_scaling}

The hyperparameters of the surrogate losses (Table~\ref{tab:hparams_surrogate}) are absolute in the units of the target: the GDI loss uses a fixed step size $\lambda$, which ties the required gradient magnitude, and hence the curvature of the functional, to the units of the target, and the energy-bound loss confines the functional between two fixed parabolas. Since the target properties differ strongly in their units and spread, the same hyperparameters would therefore strike a different balance between fitting the property and shaping the minimum of the functional for every target. To make them transferable, we multiply the labels $F^*_P$ of each target property $P$ by the factor
\begin{equation}
    k_P = \frac{\sigma_{U_0}}{\sigma_{F^*}}\,.
    \label{eq:target_scale}
\end{equation}
For extensive properties $\sigma_{F^*}$ is computed after subtracting the atomic reference (Appendix~\ref{app:output_parametrization}). All quantities in Appendix~\ref{app:output_parametrization} are computed from the scaled labels.

\newpage
\subsection{Hyperparameters}
\label{app:hyperparameters}

Tables~\ref{tab:hparams_architecture}--\ref{tab:hparams_surrogate} list the most relevant hyperparameters. All property models (geometry-only, density-input, and surrogate) as well as the direct density prediction model share the architecture and training settings of Tables~\ref{tab:hparams_architecture} and~\ref{tab:hparams_training}; the settings of Table~\ref{tab:hparams_surrogate} apply to the surrogate models only.

\begin{table}[htbp]
\centering
\caption{Architecture hyperparameters of \modelname{} and of the eSEN baseline.}
\label{tab:hparams_architecture}
\begin{tabular}{lcc}
\toprule
 & \modelname{} & eSEN \\
\midrule
Number of layers & 4 & 4 \\
Channels & 32 & 32 \\
Maximum degree $l_\text{max}$ & 4 & 4 \\
Cutoff radius (Bohr) & 6.0 & 6.0 \\
Gaussian radial basis functions & 32 & 128 \\
Edge channels & 32 & 32 \\
Hidden channels of the message MLP & 64 & 64 \\
Residual connection & post-norm & pre-norm \\
Normalization & L2 norm & RMS norm \\
Activation & Gram & gate \\
Gram activation: gating channels $N_\text{gate}$ & 16 & -- \\
Gram activation: hidden channels of the gating MLP & 512 & -- \\
Hidden channels of the readout MLP & 1024, 512 & 64, 64  \\
LayerScale initialization & 0.1 & -- \\ \midrule
Number of parameters & 6.9\,M & 2.7\,M \\
\bottomrule
\end{tabular}
\end{table}

\begin{table}[htbp]
\centering
\caption{Training hyperparameters, shared by all models.}
\label{tab:hparams_training}
\begin{tabular}{ll}
\toprule
Optimizer & AdamW, $(\beta_1, \beta_2) = (0.95, 0.99)$ \\   
Learning rate & $10^{-3}$, cosine annealing to 0 \\
Epochs & 500 \\
Batch size & 128 \\
Loss function & $L_1$ (property / density targets)\\
\bottomrule
\end{tabular}
\end{table}

\begin{table}[htbp]
\centering
\caption{\centering Hyperparameters of the surrogate training (Section~\ref{sec:surrogate_functionals}).\newline The energy bounds refer to the scaled labels (Appendix~\ref{app:target_scaling}).}
\label{tab:hparams_surrogate}
\begin{tabular}{ll}
\toprule
GDI step size $\lambda$ & 0.04 \\
GDI contraction factor $\beta$ & 0.86 \\
Energy-bound factors $a_\text{min}$, $a_\text{max}$ & 2, 5 \\
Fraction $q$ of ground-state samples & 0.5 \\
Loss weights of $\mathcal{L}_\text{GDI}$, $\mathcal{L}_\text{EB}$, $\mathcal{L}_E$, $\mathcal{L}_{\grad E}$ & 20, 1, 1, 1 \\
Cache step size $\gamma$ & adaptive, minimum $10^{-4}$ \\
Cache initialization & MINAO initial guess \\
Cache reset probability per update & 0.02 \\
\bottomrule
\end{tabular}
\end{table}

\end{document}